\documentclass{article}

\usepackage{arxiv}

\usepackage[utf8]{inputenc} % allow utf-8 input
\usepackage[T1]{fontenc}    % use 8-bit T1 fonts
\usepackage{hyperref}       % hyperlinks
\usepackage{url}            % simple URL typesetting
\usepackage{booktabs}       % professional-quality tables
\usepackage{amsfonts}       % blackboard math symbols
\usepackage{nicefrac}       % compact symbols for 1/2, etc.
\usepackage{microtype}      % microtypography
\usepackage{lipsum}		% Can be removed after putting your text content
\usepackage{graphicx}
\usepackage[square , sort, numbers]{natbib}
\usepackage{doi}
\usepackage[nolist]{acronym}
\usepackage{threeparttable} %add by song
\usepackage{tabularx} %add by song
\usepackage{multirow}
\usepackage{threeparttable}
\usepackage{makecell}
\usepackage{amsmath}
\usepackage{ragged2e}
\usepackage{capt-of}

\title{Residual Observability and Attack Detectability in Encrypted OPC UA Traffic}

\date{July 19, 2026}	% Here you can change the date presented in the paper title
\author{ \href{https://orcid.org/0000-0003-1716-5867}{\includegraphics[scale=0.06]{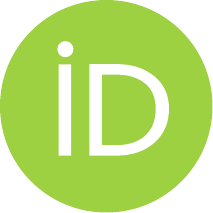}\hspace{1mm}Song Son Ha}\\
	Electrical Measurement Engineering\\
	Helmut-Schmidt-University\\
	Hamburg, Germany\\
	\texttt{song.ha@hsu-hh.de} \\
	\And
	\href{https://orcid.org/0009-0003-4437-010X}{\includegraphics[scale=0.06]{orcid.pdf}\hspace{1mm}Florian Foerster}\\
	Institute for Innovative Safety and Security\\
	Technical University of Applied Sciences Augsburg\\
	Augsburg, Germany\\
	\texttt{florian.foerster@tha.de} \\
	\And
	\href{https://orcid.org/0000-0002-5390-3946}{\includegraphics[scale=0.06]{orcid.pdf}\hspace{1mm}Henry Beuster}\\
	Electrical Measurement Engineering\\
	Helmut-Schmidt-University\\
	Hamburg, Germany\\
	\texttt{henry.beuster@hsu-hh.de} \\
	\And
	\href{https://orcid.org/0009-0006-2540-7933}{\includegraphics[scale=0.06]{orcid.pdf}\hspace{1mm}Eduard Zeller}\\
	Electrical Measurement Engineering\\
	Helmut-Schmidt-University\\
	Hamburg, Germany\\
	\texttt{eduard.zeller@hsu-hh.de} \\
	\And
	\href{https://orcid.org/0000-0003-2310-5895}{\includegraphics[scale=0.06]{orcid.pdf}\hspace{1mm}Dominik Merli}\\
	Institute for Innovative Safety and Security\\
	Technical University of Applied Sciences Augsburg\\
	Augsburg, Germany\\
	\texttt{dominik.merli@tha.de} \\
	\And
	{\hspace{1mm}Gerd Scholl}\\
	Electrical Measurement Engineering\\
	Helmut-Schmidt-University\\
	Hamburg, Germany\\
	\texttt{gerd.scholl@hsu-hh.de} \\
}

\renewcommand{\shorttitle}{Residual Observability and Attack Detectability in Encrypted OPC UA Traffic}

\hypersetup{
	hidelinks,
	pdftitle={Residual Observability and Attack Detectability in Encrypted OPC UA Traffic},
	pdfsubject={Residual structural observability and attack detectability in encrypted OPC UA traffic},
	pdfauthor={Song Son Ha; Florian Foerster; Henry Beuster; Eduard Zeller; Dominik Merli; Gerd Scholl},
	pdfkeywords={OPC UA, encrypted traffic analysis, intrusion detection, residual observability, private 5G}
}

\begin{document}
	\maketitle
	
	%\begin{acronym}
	%    \acro{iolw}[IOLW]{IO-Link Wireless}
	%    \acro{plc}[PLC]{programmable logic controller}
	%    \acro{sfrt}[SFRT]{safety function response time}
	%    \acro{ism}[ISM]{industrial, scientific and medical}
	%    \acro{mmtc}[mMTC]{massive machine type communications}
	%    \acro{iol}[IOL]{IO-Link}
	%    \acro{lte}[LTE]{long-term evolution}
	%    \acro{ue}[UE]{user equipment}
	%    \acro{nsa}[NSA]{non standalone}
	%    \acro{sa}[SA]{standalone}
	%    \acro{scs}[SCS]{subcarrier spacing}
	%    \acro{ofdm}[OFDM]{orthogonal frequency-division multiplexing}
	%    \acro{urllc}[URLLC]{ultra reliable low latency communications}
	%    \acro{embb}[eMBB]{enhanced mobile broadband}
	%    \acro{revpi}[RevPi]{Revolution Pi}
	%    \acro{vpn}[VPN]{virtual private network}
	%    \acro{dhcp}[DHCP]{Dynamic Host Configuration Protocol}
	%    \acro{ip}[IP]{Internet Protocol}
	%    \acro{rssi}[RSSI]{Received Signal Strength Indicator}
	%    \acro{iols}[IOLS]{IO-Link Safety}
	%    \acro{cpfen}[CPFEN]{Cyber Physical Finite Element Sensor Network}
	%    \acro{wmaster}[W-Master]{Wireless-Master}
	%\end{acronym}
	
\begingroup
\renewcommand{\thefootnote}{}
\footnotetext{This is the authors' version of a paper that has been accepted for presentation at the 52nd Annual Conference of the IEEE Industrial Electronics Society (IECON 2026), to be held in Doha, Qatar, on October 18--21, 2026.}
\addtocounter{footnote}{-1}
\endgroup
	
\begin{abstract}
	
	OPC Unified Architecture (OPC~UA) encryption conceals application-layer semantics and restricts intrusion detection to residual communication structure. Although machine learning-based intrusion detection systems (IDSs) can detect attacks in encrypted OPC~UA traffic, the relationship between residual structural observability and attack detectability remains insufficiently understood. This paper presents an explanatory framework combining a structural observability profile, the Structural Leakage Score (SLS), controlled within-family and cross-family comparisons, phase-specific analysis, and dimension-ablation analysis. Jensen--Shannon divergence is used to characterize transport, temporal, and protocol-lifecycle dimensions, while the SLS summarizes the residual structural magnitude. Evaluation on an industrial private~5G testbed covers four attack families with progressively reduced nominal activity. SLS generally tracks within-family recall trends but does not reproduce cross-family detectability ordering. Interpreting these mismatches also requires temporal prevalence, inter-burst persistence, predictive utility, unique contribution, and redundancy. The framework complements conventional IDS metrics by relating detection outcomes to the magnitude, temporal distribution, and predictive role of observable structural evidence.
	
\end{abstract}

\section{Introduction}
\label{sec:introduction}

OPC Unified Architecture (OPC~UA) is increasingly adopted in Industry~4.0 and industrial automation \cite{zhang23}, together with its \texttt{SignAndEncrypt} security mode for protecting application-layer semantics and sensitive operational data. Although encryption limits payload-based intrusion detection \cite{wang2022}, externally observable traffic characteristics can still support machine learning (ML)-based intrusion detection, as demonstrated for general encrypted traffic and, in initial studies, for encrypted OPC~UA communication \cite{anderson2017,neu2019}. Existing work has primarily evaluated feature representations, classifiers, and detection performance, while the relationship between residual structural observability and attack detectability remains insufficiently understood. OPC~UA provides a relevant setting for examining this relationship because connection and \texttt{SecureChannel} framing retain observable message types, lengths, and channel-management activity despite encryption \cite{OPC10000-6}. Encrypted OPC~UA traffic therefore exposes transport, temporal, and protocol-lifecycle characteristics that may be affected differently by different attack mechanisms. Accordingly, the study examines whether residual structural magnitude explains detectability trends within the same attack mechanism and whether this relationship transfers across heterogeneous attack mechanisms. This paper addresses these questions through an explanatory framework. The framework combines a structural observability profile with the Structural Leakage Score (SLS) as a first-order summary of residual structural magnitude. Controlled comparisons across configuration levels within each attack family and across attack families examine the scope of this relationship. Phase-specific analysis evaluates temporal prevalence and inter-burst persistence, whereas dimension-ablation analysis assesses predictive utility, unique contribution, and redundancy. The framework is evaluated using encrypted OPC~UA traffic collected from an industrial private~5G testbed, covering four attack families with progressively reduced nominal activity. 

The remainder of this paper is organized as follows. Section~\ref{sec:related_work} reviews related work. Sections~\ref{sec:methodology} and~\ref{sec:Experimental_setup} present the methodology and experimental setup, respectively. Sections~\ref{sec:results} and~\ref{sec:discussion} report and discuss the evaluation results, while Section~\ref{sec:conclusion} concludes the paper and outlines future work.

\section{Related Work}
\label{sec:related_work}

Encrypted-traffic analysis relies on externally observable characteristics such as packet sizes, directions, timing, flow statistics, and protocol metadata, which can support traffic classification and malicious-traffic detection despite payload confidentiality \cite{wang2022,anderson2017,papadogiannaki2021survey}. Related approaches have also been investigated in industrial environments, including encrypted ICS monitoring and DNP3 traffic analysis \cite{specht2024,detoledo2019}. For OPC~UA, Neu et al.~\cite{neu2019} showed that selected denial-of-service attacks under \texttt{SignAndEncrypt} remain detectable from observable message types and counts. ML-based intrusion detection has also been evaluated for unencrypted OPC~UA communication in industrial private~5G environments \cite{ha2026experimental}. Together, these studies show that selected OPC~UA attacks remain detectable with and without encryption, but primarily assess features, classifiers, and detection performance rather than the relationship between residual structural observability and attack detectability \cite{neu2019,ha2026experimental}.

A complementary research direction uses feature importance and explainable artificial intelligence to analyze how trained IDS models use their inputs \cite{10136827,wang2020,khan2022}. Methods such as SHAP and permutation importance identify features that influence model predictions or contribute to detection performance \cite{lundberg2017,fisher2019,SARHAN2022100359}. However, these analyses remain tied to a fitted classifier, its predictions, or its decision function \cite{tritscher2023}. They do not independently characterize residual structural magnitude, its dimensional composition, or its temporal distribution within an attack campaign.

Existing research therefore addresses either whether OPC~UA attacks can be detected or how trained models use observable features. The relationship between residual structural observability and attack detectability remains less explored. In particular, it is unclear when residual structural magnitude explains detectability trends within the same attack mechanism, whether this relationship transfers across heterogeneous attack mechanisms, and how temporal prevalence, inter-burst persistence, predictive utility, and redundancy influence this relationship. This work addresses this gap through controlled within-family and cross-family comparisons, phase-specific analysis, and dimension-ablation analysis.

\section{Methodology}
\label{sec:methodology}

\begin{table}[t]
	\centering
	\caption{Observable features used for IDS training and residual structural observability analysis.}
	\label{tab:structure_mapping_full}
	\small
	\renewcommand{\arraystretch}{1.2}
	\begin{tabularx}{\textwidth}{
			@{}
			>{\raggedright\arraybackslash}p{0.10\textwidth}
			@{\hspace{6pt}}
			>{\raggedright\arraybackslash}p{0.21\textwidth}
			@{\hspace{6pt}}
			>{\raggedright\arraybackslash}p{0.37\textwidth}
			@{\hspace{6pt}}
			>{\raggedright\arraybackslash}X
			@{}
		}
		\toprule
		\textbf{Dimension} &
		\textbf{Feature Group} &
		\textbf{Features} &
		\textbf{Description} \\
		\midrule
		
		\multirow{5}{*}{\textbf{Transport}}
		& Volume
		& \texttt{total\_\{pkt, bytes, body\_\{bytes, count\}\}}
		& Aggregate packet, byte, and message-body volume. \\
		
		& Directionality
		& \texttt{total\_\{fwd, bwd\}\_\{pkt, bytes\}}
		& Packet and byte volume in both directions. \\
		
		& Asymmetry
		& \texttt{fwd\_minus\_bwd\_\{pkt, byte\}\_norm}
		& Normalized directional packet and byte imbalance. \\
		
		& Packet Characteristics
		& \texttt{mean\_pkt\_size}, \texttt{body\_bytes\_ratio}
		& Average packet size and message-body byte ratio. \\
		
		& Flow Dominance
		& \texttt{max\_flow\_pkt\_ratio}
		& Traffic concentration in the most active flow. \\
		
		\midrule
		
		\multirow{4}{*}{\textbf{Temporal}}
		& Flow Dynamics
		& \texttt{delta\_pkt},
		\texttt{flow\_count},
		\texttt{flow\_churn}
		& Changes in packet volume and active flows. \\
		
		& Flow Distribution
		& \texttt{flow\_pkt\_entropy}
		& Packet distribution across active flows. \\
		
		& Rate and Timing
		& \texttt{pkt\_rate},
		\texttt{\{mean, total\}\_iat},
		\texttt{total\_iat\_count}
		& Packet rate and inter-arrival timing. \\
		
		& Duration and Variability
		& \texttt{\{mean, max\}\_flow\_duration},
		\texttt{std\_flow\_pkt\_count},
		\texttt{burstiness}
		& Flow duration, packet-count variability, and burstiness. \\
		
		\midrule
		
		\multirow{3}{*}{\makecell{\textbf{Protocol-}\\\textbf{lifecycle}}}
		& Connection Activity
		& \texttt{total\_\{handshake, hel, ack, err\}}
		& Connection establishment and error activity. \\
		
		& Secure Channel Activity
		& \texttt{total\_\{opn, clo, secch\}},
		\texttt{sec\_ch\_churn}
		& \texttt{SecureChannel} establishment, closure, and churn. \\
		
		& Handshake Dynamics
		& \texttt{handshake\_rate\_pkt},
		\texttt{delta\_handshake},
		\texttt{handshake\_ratio\_delta}
		& Handshake frequency and temporal variation. \\
		
		\bottomrule
	\end{tabularx}
\end{table}

\subsection{Observable Features and Structural Dimensions}
\label{subsec:observable_dimensions}

Under OPC~UA encryption, application-layer service semantics are concealed, but residual communication structure remains observable. Encrypted traffic is segmented into consecutive, non-overlapping 5\,s time windows (TWs). Within each TW, bidirectional flows are processed independently, and payload-content-independent traffic and protocol-framing statistics are extracted. The resulting flow-level statistics are aggregated across all active flows to form one feature vector per TW. The extracted features are organized into transport, temporal, and protocol-lifecycle dimensions. The transport dimension captures traffic volume, packet-size characteristics, directionality, and flow distribution, while the temporal dimension represents communication rates, timing, duration, variability, and changes in flow activity. The protocol-lifecycle dimension captures observable OPC~UA connection-establishment and \texttt{SecureChannel} management activity. Although \texttt{SignAndEncrypt} protects service bodies, outer framing remains identifiable, including \texttt{HEL}, \texttt{ACK}, \texttt{ERR}, \texttt{OPN}, \texttt{MSG}, and \texttt{CLO} message types \cite{OPC10000-6}.

Table~\ref{tab:structure_mapping_full} summarizes the extracted features and their assignment to the three structural dimensions. The same feature representation is used for both IDS training and residual structural observability analysis.

\subsection{Structural Observability Profile and SLS}
\label{subsec:sls_formulation}

The structural observability profile characterizes residual structural deviation across transport, temporal, and protocol-lifecycle dimensions. SLS then provides a model-agnostic, first-order summary of their overall residual structural magnitude between benign and attack-labeled encrypted OPC~UA traffic. For each evaluation split, let $\mathrm{B}$ denote the windows from benign-only recordings. For attack family $a$ and configuration level $\ell$, let $\mathrm{C}_{a,\ell}$ denote all attack-labeled windows from the first to the last attack-burst window, including inter-burst windows.
Let $
\mathcal{F}
=
\left\{
\mathcal{F}_{\mathrm{Trans}},
\mathcal{F}_{\mathrm{Temp}},
\mathcal{F}_{\mathrm{Prot}}
\right\}
$ 
denote the transport, temporal, and protocol-lifecycle feature subsets. Non-finite values are imputed using per-feature medians fitted on benign training windows. For each feature $j$, the benign distribution $P_j$ and attack-labeled distribution $Q_{j,a,\ell}$ are estimated using fixed quantile-bin boundaries fitted once on the imputed benign training distribution. A target of 30 bins is used, duplicate boundaries are removed, and the outer bins extend to $-\infty$ and $+\infty$. The resulting feature-specific bins $K_j\leq30$ remain fixed for all subsequent comparisons.

For bin $k\in\{1,\ldots,K_j\}$, the corresponding empirical probabilities are

\begin{equation}
	p_{j,k} = \frac{n^{\mathrm{B}}_{j,k}+\epsilon}{\sum_{r=1}^{K_j}n^{\mathrm{B}}_{j,r}+K_j\epsilon},
	\quad
	q_{j,k,a,\ell} = \frac{n^{\mathrm{C}}_{j,k,a,\ell}+\epsilon}{\sum_{r=1}^{K_j}n^{\mathrm{C}}_{j,r,a,\ell}+K_j\epsilon},
\end{equation}

where $n^{\mathrm{B}}_{j,k}$ and $n^{\mathrm{C}}_{j,k,a,\ell}$ denote the numbers of windows in $\mathrm{B}$ and $\mathrm{C}_{a,\ell}$ assigned to bin $k$, respectively, and $\epsilon=10^{-12}$ is a pseudocount added to each bin. These probabilities define the discrete distributions $P_j=(p_{j,1},\ldots,p_{j,K_j})$ and $Q_{j,a,\ell}=(q_{j,1,a,\ell},\ldots,q_{j,K_j,a,\ell})$.

Their equally weighted mixture distribution is defined as

\begin{equation}
	M_{j,a,\ell}
	=
	\frac{1}{2}
	\left(
	P_j + Q_{j,a,\ell}
	\right).
\end{equation}

The feature-level residual structural deviation is then measured using the Jensen--Shannon divergence (JSD):

\begin{equation}
	\operatorname{JSD}(P_j, Q_{j,a,\ell})
	=
	\frac{
		D_{\mathrm{KL}}(P_j \| M_{j,a,\ell})
		+
		D_{\mathrm{KL}}(Q_{j,a,\ell} \| M_{j,a,\ell})
	}{2},
\end{equation}

where $D_{\mathrm{KL}}$ denotes the Kullback--Leibler divergence.

Using base-two logarithms, JSD is symmetric and bounded within $[0,1]$, with larger values indicating stronger marginal deviation from benign traffic.

The divergence within structural dimension $i$ is obtained by averaging the feature-level residual structural deviation:

\begin{equation}
	D_i(a,\ell)
	=
	\frac{1}{|\mathcal{F}_i|}
	\sum_{j\in\mathcal{F}_i}
	\operatorname{JSD}
	\left(
	P_j,Q_{j,a,\ell}
	\right).
\end{equation}

The overall SLS is obtained by averaging the three dimension-level divergences:

\begin{equation}
	\operatorname{SLS}(a,\ell)
	=
	\frac{
		D_{\mathrm{Trans}}(a,\ell)
		+
		D_{\mathrm{Temp}}(a,\ell)
		+
		D_{\mathrm{Prot}}(a,\ell)
	}{3}.
\end{equation}

This gives all dimensions equal weight, regardless of feature count. Because SLS aggregates feature-wise divergences, it summarizes marginal residual structural magnitude but does not capture how multiple features jointly separate benign and attack-labeled windows. Consequently, SLS is intended as a computationally efficient, first-order diagnostic baseline rather than a complex multivariate predictor.

\begin{table}[b]
	\centering
	\caption{Evaluated attack families and configuration progression from the highest nominal activity at L1 to the lowest at L4.}
	\label{tab:attack_configs}
	\small
	\renewcommand{\arraystretch}{1.10}
	\setlength{\tabcolsep}{5pt}
	
	\begin{tabularx}{\textwidth}{
			@{}
			>{\raggedright\arraybackslash}p{0.13\textwidth}
			@{\hspace{-6pt}}
			>{\justifying\setlength{\parindent}{0pt}\arraybackslash}p{0.35\textwidth}
			@{\hspace{10pt}}
			>{\justifying\setlength{\parindent}{0pt}\arraybackslash}X
			@{}
		}

		\toprule
		\textbf{Family}
		&
		\textbf{Mechanism}
		&
		\textbf{Configuration Progression from L1 to L4}
		\\
		\midrule
		
		BrowseAS
		&
		Performs bounded random walks through the OPC~UA address space using \texttt{Browse} service requests over a pre-discovered graph.
		&
		The traversal-initiation rate, concurrent \texttt{Sessions}, walk depth, and requested references per node are progressively reduced. Active bursts are shortened, idle intervals extended, and \texttt{Sessions} recreated less frequently.
		\\
		\midrule
		
		PubFlood
		&
		Creates one \texttt{Subscription} per \texttt{Session} and submits batches of \texttt{Publish} service requests without waiting for responses.
		&
		The submission-cycle rate, number of concurrent \texttt{Sessions}, and number of \texttt{Publish} service requests per cycle are progressively reduced. Active bursts are shortened and idle intervals are extended.
		\\
		\midrule
		
		\makecell[tl]{PSC\\Exhaustion}
		&
		Repeatedly establishes and concurrently retains \texttt{SecureChannels} with associated \texttt{Sessions}, using periodic \texttt{Read} service requests to maintain activity on the retained connections.
		&
		The establishment-attempt rate, number of parallel client instances, and retained \texttt{SecureChannel}--\texttt{Session} pairs per instance are progressively reduced. The retention period is increased, while burst--idle timing and periodic \texttt{Read} activity are adjusted for less frequent operation.
		\\
		\midrule
		
		TranslateBP
		&
		Issues \texttt{TranslateBrowsePathsToNodeIds} service requests containing multiple distinct \texttt{RelativePaths} of configurable depth.
		&
		The per-\texttt{Session} request rate, number of concurrent \texttt{Sessions}, \texttt{RelativePaths} per request, and \texttt{RelativePath} depth are progressively reduced. Active bursts are shortened and idle intervals extended.
		\\
		
		\bottomrule
	\end{tabularx}
	
\end{table}

\subsection{Attack Design}
\label{subsec:attacks}

Four protocol-aware OPC~UA attack families are evaluated under \texttt{SignAndEncrypt}: Browse Address Space (BrowseAS), Publish Request Flooding (PubFlood), Persistent Secure Channel Exhaustion (PSC Exhaustion), and Translate Browse Path (TranslateBP). The attacks are adapted from Claroty's OPC~UA exploitation framework \cite{claroty2023} and remain syntactically valid during execution. The attacker is assumed to control an OPC~UA client whose certificate is already trusted by the server. Certificate acquisition and trust-list compromise are outside the scope of this work. The selected attack families cover different sources of residual structural observability. BrowseAS, PubFlood, and TranslateBP manipulate OPC~UA service-request workloads and primarily affect transport and temporal characteristics. PSC Exhaustion additionally manipulates the establishment and retention of \texttt{SecureChannels} and associated sessions, thereby affecting the protocol-lifecycle dimension. For each attack family, four configuration levels, denoted L1 through L4, progressively reduce nominal activity while preserving the underlying attack mechanism, as summarized in Table~\ref{tab:attack_configs}. Depending on the family, the progression reduces the traversal-initiation rate, submission-cycle rate, establishment-attempt rate, or service-request rate, together with concurrency or workload complexity. It also adjusts burst and idle timing and other lifecycle-related timing parameters. These controlled within-family variations enable the relationship between nominal activity, residual structural magnitude, and attack detectability to be examined without changing the underlying attack mechanism.

\section{Experimental Setup}
\label{sec:Experimental_setup}

\subsection{Industrial Private~5G Testbed}
\label{subsec:testbed}

The evaluation is conducted on the industrial private~5G testbed initially reported in~\cite{11205743} and extended in~\cite{ha2026experimental}. The testbed comprises a 3GPP Release~16 Standalone network, heterogeneous industrial endpoints, an OPC~UA-based warehouse automation environment, and an IDS backend. The present study extends this setup to support encrypted OPC~UA communication and payload-independent traffic analysis.

Industrial endpoints include Revolution~Pi controllers, Raspberry~Pi~5 devices, and Intel NUC systems connected through 5G communication modules. Engineering, diagnostic, and attack nodes are integrated through industrial routers, while attacks are executed from a dedicated workstation. The warehouse automation environment includes PLC-controlled conveyor systems, sensors, and actuators simulated using Factory~I/O~\cite{factoryio}. Process data is exchanged between the simulated factory environment and the controllers, while OPC~UA servers and clients running on controllers expose and access process variables over the private~5G network. The resulting communication covers HMI supervision, SCADA monitoring, engineering access, and remote PLC control. All OPC~UA communication operates under \texttt{SignAndEncrypt} using the \texttt{Aes128-Sha256-RsaOaep} security policy. Mirrored traffic is processed on a backend system for feature extraction, TW aggregation, and dataset generation. The R\&S~PACE~2 deep packet inspection library is used to extract protocol-aware features without access to decrypted OPC~UA service content.

\subsection{Dataset Construction}
\label{subsec:dataset_anomaly_detection}

The dataset collected from the private 5G testbed comprises 250 packet capture (PCAP) recordings of 10\,min each, totaling approximately 42\,h of encrypted traffic. Normal operations contribute 90 benign recordings, partitioned into 63 training, 9 validation, and 18 test PCAPs across various workloads. The remaining 160 recordings span four attack families and four configuration levels, providing 10 PCAPs per combination. This attack traffic is divided into 80 training, 32 validation, and 48 test PCAPs. Windows from benign-only recordings are assigned the benign label. For each attack recording, the attack campaign extends from the first to the last attack-burst window, and all windows within this interval are assigned the attack label. The attack-labeled windows comprise both attack-burst and inter-burst windows. The latter are not treated as benign because communication effects may persist between successive bursts. This labeling strategy reflects operational realism by retaining inter-burst windows, during which attackers may maintain cryptographic connections despite reduced or absent attack activity. Windows outside the attack campaign are excluded from model development and evaluation.

\subsection{Detectability and Structural Evaluation}
\label{subsec:evaluation}

Attack detectability is evaluated through supervised binary classification of benign and attack-labeled windows using the extracted feature vectors. Four ML-based IDS models are considered: Logistic Regression (LogReg), Random Forest (RF), Support Vector Machine (SVM) with a radial basis function kernel, and Extreme Gradient Boosting (XGBoost).

Overall IDS performance is measured using precision, recall, F1-score, the area under the receiver operating characteristic curve (AUROC), the area under the precision-recall curve (AUPRC), and false-positive rate (FPR). For each model, the decision threshold is selected on the validation set by maximizing the attack-class F1-score while limiting the FPR on benign validation windows to at most 5\,\%. The selected threshold is then kept fixed for evaluation on the held-out test set. Recall is used as the primary threshold-dependent measure of attack detectability because it directly reflects the proportion of attack-labeled windows detected by the IDS. To reduce dependence on a particular classifier, results are summarized using the median recall across the four models.

Residual structural observability and attack detectability are compared across configuration levels L1 through L4 within each attack family and across attack families using pooled estimates from all four levels. For the level-wise and pooled SLS estimates, 95\,\% confidence intervals (CIs) are obtained from 1,000 PCAP-level bootstrap repetitions using the 2.5th and 97.5th percentiles. Benign and attack PCAPs are resampled with replacement while the training-fitted medians and bin boundaries remain fixed. For pooled estimates, attack PCAPs are resampled separately within each configuration level. Phase-specific analysis examines temporal prevalence and inter-burst persistence. Dimension-ablation analysis evaluates the predictive utility, unique contribution, and redundancy of each structural dimension using single-dimension and dimension-removal feature sets.

\section{Evaluation Results}
\label{sec:results}

\subsection{Baseline IDS Performance}
\label{subsec:baseline_ids_performance}

Table~\ref{tab:baseline_ids_performance} summarizes the overall performance of the four ML-based IDS models on the held-out test set. All models achieve precision close to 0.98, recall above 0.81, and F1-scores between approximately 0.89 and 0.90. Their AUROC and AUPRC values also remain consistently high, at approximately 0.93 and 0.97, respectively. The test-set FPR remains close to the 5\,\% constraint imposed during threshold selection. Although XGBoost yields the highest recall and RF the lowest FPR, performance variations across models remain marginal.

Furthermore, Figure~\ref{fig:family_recall} shows a consistent family-level ordering across the four classifiers. PubFlood achieves the highest recall, followed by BrowseAS and TranslateBP, whereas PSC Exhaustion is substantially less detectable. The consistent ordering across the four models indicates that the observed family-level differences are not specific to a single classifier.

\begin{table}[!h]
	\centering
	
	\caption{Overall IDS performance on the held-out test set.}
	\label{tab:baseline_ids_performance}
	
	{
		\small
		\renewcommand{\arraystretch}{1.08}
		\setlength{\tabcolsep}{5pt}
		
		\begin{tabular}{@{}lrrrrrr@{}}
			\toprule
			\textbf{Model} &
			\textbf{Precision} &
			\textbf{Recall} &
			\textbf{F1-score} &
			\textbf{AUROC} &
			\textbf{AUPRC} &
			\textbf{FPR} \\
			\midrule
			LogReg & 0.975 & 0.813 & 0.887 & 0.930 & 0.974 & 0.050 \\
			RF       & \textbf{0.977} & 0.826 & 0.895 & 0.932 & 0.975 & \textbf{0.046} \\
			SVM                 & 0.976 & 0.821 & 0.892 & 0.931 & 0.974 & 0.049 \\
			XGBoost             & 0.976 & \textbf{0.839} & \textbf{0.903} & \textbf{0.935} & \textbf{0.976} & 0.050 \\
			\bottomrule
		\end{tabular}
	}
	
	\vspace{1em}
	
	\begin{minipage}{\linewidth}
		\centering
		\includegraphics[width=0.7\linewidth]
		{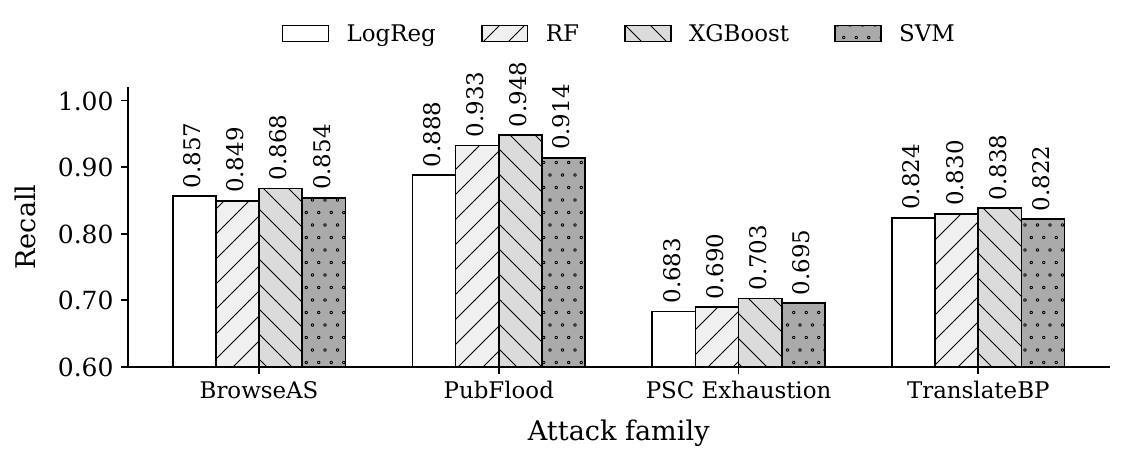}
		\captionof{figure}{Recall by attack family, pooled across all four configuration levels.}
		\label{fig:family_recall}
	\end{minipage}
	
\end{table}

\begin{table}[t]
	\centering
	\caption{Level-wise decomposition and residual structural observability across attack configurations.}
	\label{tab:sls_levelwise_decomposition}
	
	\small
	\renewcommand{\arraystretch}{1.08}
	\setlength{\tabcolsep}{4pt}
	
	\begin{tabular}{l c|ccc|c}
		\toprule
		\textbf{Attack} &
		\textbf{Config.} &
		\textbf{D$_{\mathbf{Trans}}$} &
		\textbf{D$_{\mathbf{Temp}}$} &
		\textbf{D$_{\mathbf{Prot}}$} &
		\textbf{SLS} \\
		\midrule
		
		\multirow{4}{*}{BrowseAS}
		& L1 & \textbf{0.645} & 0.374 & 0.017 & 0.345 \\
		& L2 & \textbf{0.474} & 0.238 & 0.007 & 0.240 \\
		& L3 & \textbf{0.108} & 0.093 & 0.010 & 0.070 \\
		& L4 & 0.034 & \textbf{0.055} & 0.008 & 0.032 \\
		
		\midrule
		
		\multirow{4}{*}{PubFlood}
		& L1 & \textbf{0.266} & 0.178 & 0.012 & 0.152 \\
		& L2 & \textbf{0.200} & 0.119 & 0.019 & 0.113 \\
		& L3 & \textbf{0.127} & 0.125 & 0.016 & 0.090 \\
		& L4 & \textbf{0.089} & 0.065 & 0.029 & 0.061 \\
		
		\midrule
		
		\multirow{4}{*}{\makecell{PSC\\Exhaustion}}
		& L1 & \textbf{0.344} & 0.171 & 0.293 & 0.269 \\
		& L2 & \textbf{0.152} & 0.086 & 0.115 & 0.118 \\
		& L3 & 0.120 & 0.084 & \textbf{0.127} & 0.110 \\
		& L4 & \textbf{0.062} & 0.044 & 0.055 & 0.054 \\
		
		\midrule
		
		\multirow{4}{*}{TranslateBP}
		& L1 & \textbf{0.375} & 0.167 & 0.010 & 0.184 \\
		& L2 & \textbf{0.313} & 0.098 & 0.023 & 0.145 \\
		& L3 & \textbf{0.140} & 0.087 & 0.028 & 0.085 \\
		& L4 & 0.070 & \textbf{0.074} & 0.015 & 0.053 \\
		
		\bottomrule
	\end{tabular}
\end{table}

\subsection{Residual Structural Observability}

Table~\ref{tab:sls_levelwise_decomposition} reports the structural observability profile and SLS for each attack family and configuration level. SLS decreases monotonically from L1 to L4 across all four attack families, showing that reduced nominal activity is accompanied by lower residual structural magnitude. BrowseAS exhibits the largest overall reduction, with a particularly sharp decrease between L2 and L3, whereas PubFlood declines more gradually across the four levels. The dimensional composition also differs across attack mechanisms. BrowseAS, PubFlood, and TranslateBP exhibit primarily transport and temporal divergence, with limited protocol-lifecycle contribution. PSC Exhaustion shows a distinct profile with substantial protocol-lifecycle divergence, which slightly exceeds the transport divergence at L3.

\begin{figure}[b]
	\centering
	\includegraphics[width=0.7\linewidth]{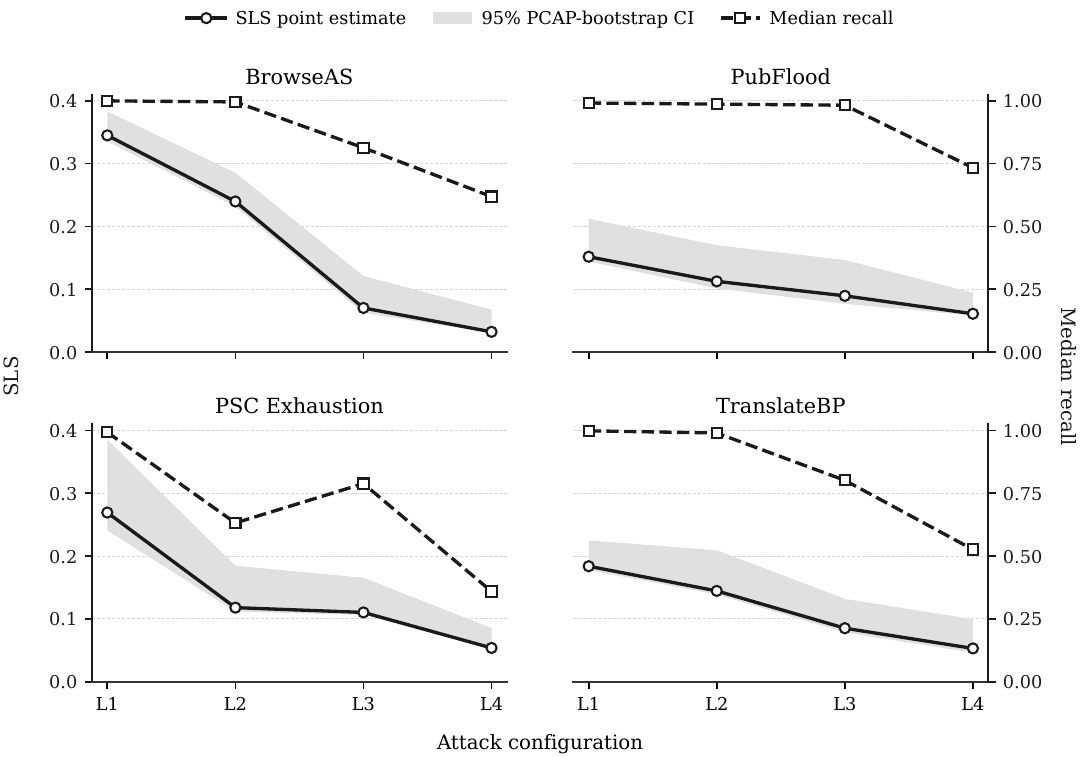}
	\caption{SLS and median recall across configuration levels and attack families.}
	\label{fig:figure_level_aligned_sls_recall_dual_axis}
\end{figure}

\subsection{Within-Family Structural Observability and Detectability}

Figure~\ref{fig:figure_level_aligned_sls_recall_dual_axis} compares SLS and median recall across configuration levels within each attack family. For all four families, both measures reach their lowest values at L4, showing that the lowest nominal activity is associated with lower residual structural magnitude and attack detectability.

For BrowseAS, SLS decreases from 0.345 at L1 to 0.032 at L4, while median recall decreases from 1.000 to 0.619. PubFlood exhibits a more gradual SLS decline, whereas recall remains close to 1.0 through L3 before falling to 0.733 at L4. TranslateBP shows a similar overall relationship, with recall remaining high at L1 and L2 before declining more clearly at L3 and L4. These results show that lower residual structural magnitude is generally accompanied by lower attack detectability within the same attack mechanism, although the changes are not proportional. PSC Exhaustion follows the same overall decline from L1 to L4 but shows a different ordering at L2 and L3. SLS decreases slightly from 0.118 to 0.110, while median recall increases from 0.632 to 0.788. Thus, SLS captures the general within-family decline but does not explain the detectability ordering between every adjacent configuration level. The following cross-family analysis examines whether the overall within-family relationship transfers across heterogeneous attack mechanisms.

\begin{table}[b]
	\centering
	\caption{Attack-burst fraction (\%) among attack-labeled windows.}
	\label{tab:attack_burst_fraction}
	
	\small
	
	\renewcommand{\arraystretch}{1.08}
	\setlength{\tabcolsep}{6pt}
	
	\begin{tabular}{lrrrr}
		\toprule
		\textbf{Attack Family} &
		\textbf{L1} &
		\textbf{L2} &
		\textbf{L3} &
		\textbf{L4} \\
		\midrule
		
		BrowseAS
		& 100.0 & 94.2 & 55.7 & 38.3 \\
		
		PubFlood
		& 93.7 & 89.6 & 64.9 & 65.2 \\
		
		PSC Exhaustion
		& 92.2 & 54.8 & 49.9 & 21.2 \\
		
		TranslateBP
		& 89.6 & 87.5 & 49.9 & 24.6 \\
		
		\bottomrule
	\end{tabular}
	
\end{table}

\subsection{Cross-Family Structural Observability and Detectability}

\begin{figure}[t]
	\centering
	\includegraphics[width=0.6\linewidth]{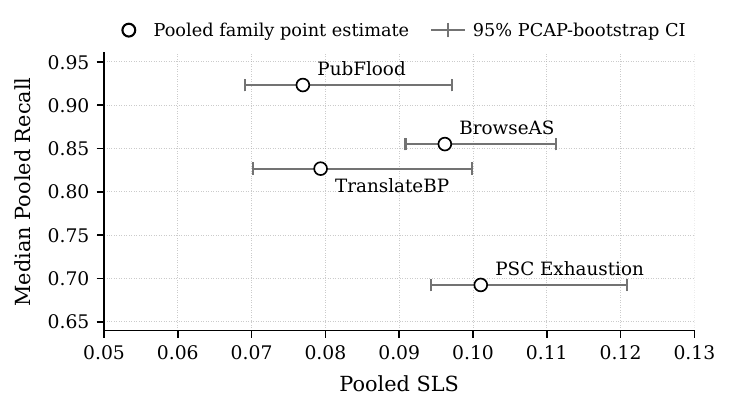}
	\caption{Cross-family relationship between pooled SLS and median recall.}
	\label{fig:figure_cross_family_sls_vs_median_recall}
\end{figure}

Figure~\ref{fig:figure_cross_family_sls_vs_median_recall} compares pooled SLS and median recall across attack families. PSC Exhaustion has the highest pooled SLS at 0.101, followed by BrowseAS at 0.096, TranslateBP at 0.079, and PubFlood at 0.077. Detectability follows a different ordering. PubFlood achieves the highest median recall at 0.923, followed by BrowseAS at 0.855 and TranslateBP at 0.827, whereas PSC Exhaustion has the lowest recall at 0.693. Several SLS confidence intervals also overlap. The pooled results show that the within-family relationship between residual structural magnitude and attack detectability does not transfer directly across heterogeneous attack mechanisms. In particular, the highest residual structural magnitude coincides with the lowest recall for PSC Exhaustion, while PubFlood exhibits the opposite pattern. The following phase-specific analysis examines whether temporal prevalence and inter-burst persistence could explain part of this mismatch.

\subsection{Temporal Evidence Prevalence and Persistence}
\label{subsec:temporal_evidence}

Residual structural magnitude does not show how observable structural evidence is distributed over an attack campaign. Phase-specific analysis therefore separates attack-labeled windows into attack-burst and inter-burst windows without changing their binary labels. Campaign-level results remain based on all attack-labeled windows. Temporal persistence describes the extent to which residual structural observability and attack detectability remain during inter-burst windows.

\subsubsection{Temporal Prevalence of Attack Bursts}

Temporal prevalence is measured as the fraction of attack-labeled windows categorized as attack-burst windows. As shown in Table~\ref{tab:attack_burst_fraction}, this fraction generally decreases as nominal activity is reduced. PSC Exhaustion shows the largest overall decline, from 92.2\,\% at L1 to 21.2\,\% at L4. BrowseAS and TranslateBP decrease to 38.3\,\% and 24.6\,\%, respectively, while PubFlood retains an attack-burst fraction of 65.2\,\% at L4. The configuration levels therefore differ not only in nominal activity, but also in the proportion of the attack campaign occupied by attack-burst windows.

\begin{figure}[t]
	\centering
	\includegraphics[width=0.8\linewidth]
	{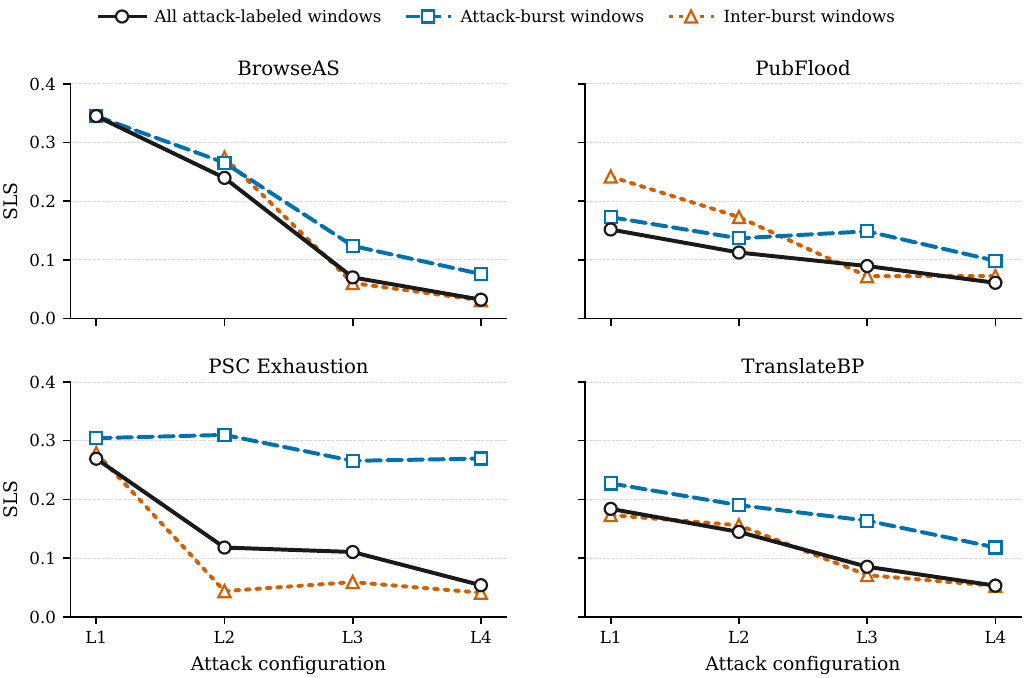}
	\caption{SLS for campaign, attack-burst, and inter-burst windows.}
	\label{fig:phase_specific_sls}
\end{figure}

\begin{figure}[t]
	\centering
	\includegraphics[width=0.8\linewidth]
	{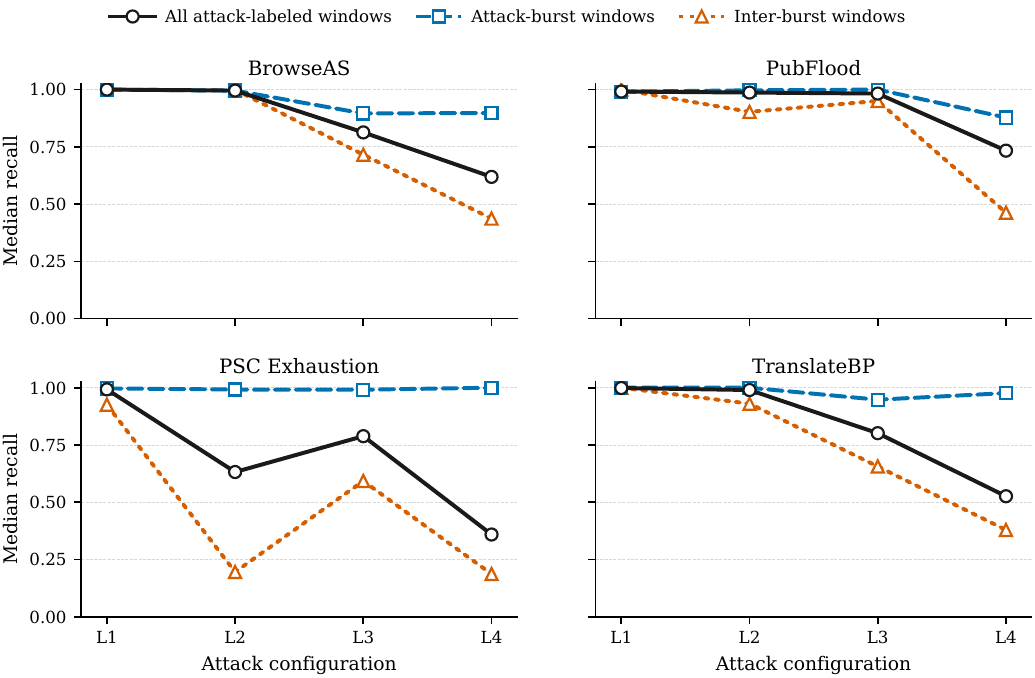}
	\caption{Median recall for campaign, attack-burst, and inter-burst windows.}
	\label{fig:phase_specific_recall}
\end{figure}

\subsubsection{Persistence of Observable Structural Evidence}

Temporal persistence is assessed by comparing SLS for attack-burst and inter-burst windows. For each combination of attack family and configuration level, windows from the two phases are pooled separately across the corresponding test PCAPs, while campaign-level SLS is computed from all attack-labeled windows. Because JSD is computed after pooling, campaign-level SLS is not a weighted average of the two phase-specific values.

As shown in Table~\ref{tab:attack_burst_fraction} and Fig.~\ref{fig:phase_specific_sls}, BrowseAS has no inter-burst SLS value at L1 because all attack-labeled windows are attack-burst windows. Its phase-specific SLS values remain similar at L2, while attack-burst SLS becomes clearly higher at L3 and L4 as inter-burst residual structural magnitude decreases. TranslateBP exhibits higher attack-burst SLS across all four levels, indicating consistently weaker residual structural magnitude during inter-burst windows. PubFlood shows a different pattern, with inter-burst SLS higher than attack-burst SLS at L1 and L2, before becoming lower at L3 and L4. This reversal indicates that residual structural magnitude is more persistent between bursts at the higher-activity levels but becomes increasingly concentrated within attack-burst windows as nominal activity is reduced. 

PSC Exhaustion shows the clearest difference between the two phases. Attack-burst SLS remains high across all levels, while inter-burst SLS is substantially lower from L2 onward. As attack-burst prevalence decreases, the weaker inter-burst evidence accounts for a larger share of the campaign-level distributions, while residual structural magnitude during attack-burst windows remains high.

\subsubsection{Detectability During and Between Bursts}

Figure~\ref{fig:phase_specific_recall} reports median recall for all attack-labeled windows and separately for attack-burst and inter-burst windows. Campaign-level recall is first computed for each classifier as the window-count-weighted combination of the two phase-specific recall values. The median is then calculated across the four models. BrowseAS and TranslateBP maintain high attack-burst recall across configuration levels, while their inter-burst recall declines more clearly at L3 and L4. Similarly, PubFlood retains high recall in both phases through L3, followed by a sharp inter-burst decline at L4. 
PSC Exhaustion shows a stronger contrast between the two phases. Attack-burst recall remains close to 1.0 across all levels, whereas inter-burst recall is substantially lower from L2 onward. Combined with the decreasing attack-burst fraction, this produces lower campaign-level recall despite consistently high detectability during attack-burst windows. The different campaign-level ordering of PSC Exhaustion at L2 and L3 originates from the inter-burst windows. Attack-burst recall is similarly high at both levels, but inter-burst recall is considerably higher at L3. The corresponding inter-burst SLS values differ only slightly and therefore do not explain the larger difference in recall.

\begin{table}[t]
	\centering
	\caption{Dimension-ablation recall for pooled attack families.}
	\label{tab:dimension_ablation}
	
	\renewcommand{\arraystretch}{1.10}
	\setlength{\tabcolsep}{5.0pt}
	
	\resizebox{0.85\textwidth}{!}{
		\begin{tabular}{lccccccc}
			\toprule
			
			\raisebox{-2.0ex}[0pt][0pt]{\textbf{Attack Family}} &
			\multicolumn{1}{c}{\textbf{Full}} &
			\multicolumn{3}{c}{\textbf{Single-dimension feature sets}} &
			\multicolumn{3}{c}{\textbf{Dimension-removal feature sets}} \\
			
			\cmidrule(lr){2-2}
			\cmidrule(lr){3-5}
			\cmidrule(lr){6-8}
			
			&
			\textbf{All} &
			\textbf{Trans. only} &
			\textbf{Temp. only} &
			\textbf{Prot. only} &
			\textbf{w/o Trans.} &
			\textbf{w/o Temp.} &
			\textbf{w/o Prot.} \\
			\midrule
			
			BrowseAS
			& 0.855
			& 0.580
			& 0.828
			& 0.117
			& 0.857
			& 0.583
			& 0.852 \\
			
			PubFlood
			& 0.923
			& 0.774
			& 0.855
			& 0.103
			& 0.894
			& 0.771
			& 0.916 \\
			
			PSC Exhaustion
			& 0.693
			& 0.516
			& 0.680
			& 0.652
			& 0.695
			& 0.668
			& 0.686 \\
			
			TranslateBP
			& 0.827
			& 0.593
			& 0.798
			& 0.053
			& 0.815
			& 0.598
			& 0.822 \\
			
			\bottomrule
		\end{tabular}
	}
\end{table}

\subsection{Dimension-Ablation Analysis}
\label{subsec:dimension_ablation}

Table~\ref{tab:dimension_ablation} reports pooled median recall after retraining classifiers across full, single-dimension, and dimension-removal feature sets. The same PCAP-level partitions are retained for all feature sets. Single-dimension results measure isolated predictive utility, while dimension-removal results assess unique contribution beyond the remaining dimensions.

The temporal dimension provides the strongest isolated predictive utility across all four attack families, with temporal-only recall ranging from 0.680 for PSC Exhaustion to 0.855 for PubFlood. Transport-only recall is lower, while the protocol-lifecycle dimension provides limited isolated predictive utility for BrowseAS, PubFlood, and TranslateBP. PSC Exhaustion differs, with protocol-lifecycle-only recall reaching 0.652, consistent with its substantial protocol-lifecycle divergence.

Removing the temporal dimension produces the largest recall reductions for BrowseAS, PubFlood, and TranslateBP. Removing either the transport or protocol-lifecycle dimension causes smaller changes. PSC Exhaustion exhibits greater redundancy across the structural dimensions. Both its temporal-only and protocol-lifecycle-only sets achieve high recall, while removing either from the full set causes minimal reduction. Thus, the protocol-lifecycle dimension provides substantial isolated predictive utility for PSC Exhaustion but limited unique contribution beyond the remaining dimensions.

\section{Discussion}
\label{sec:discussion}

The results show that residual structural observability and attack detectability are related but distinct. SLS provides a model-agnostic, first-order summary of residual structural magnitude. Within the same attack mechanism, decreasing SLS generally follows the decline in recall as nominal activity is reduced. Residual structural magnitude can therefore explain broad within-family detectability trends when the underlying attack mechanism remains unchanged.

The cross-family results establish the main limit of this relationship. PSC Exhaustion exhibits the highest pooled SLS but the lowest campaign-level recall, while PubFlood shows the opposite ordering. This mismatch occurs because attack families may differ in how observable structural evidence is distributed across dimensions and over time, even when their overall residual structural magnitude is similar. SLS alone therefore cannot rank attack families by detectability, and the within-family relationship between SLS and recall does not directly transfer across heterogeneous attack mechanisms.

PSC Exhaustion provides the clearest diagnostic case. Its attack-burst SLS remains high, its protocol-lifecycle divergence is substantial, and attack-burst recall remains close to saturation. Nevertheless, campaign-level recall decreases because attack-burst windows become less prevalent and inter-burst evidence remains weaker. Low campaign-level recall therefore does not necessarily indicate that attack-burst windows are structurally unobservable. It may instead result from strong observable structural evidence occurring in only a small part of the attack campaign. The different recall ordering at L2 and L3 is also not explained by their similar inter-burst SLS values, indicating that marginal residual structural magnitude alone does not capture the remaining differences in multivariate separability and model utilization.

The dimension-ablation results further separate residual structural observability from predictive utility. The temporal dimension provides the strongest isolated predictive utility and unique contribution for BrowseAS, PubFlood, and TranslateBP. For PSC Exhaustion, the protocol-lifecycle dimension exhibits substantial divergence and high isolated recall, but limited unique contribution. A structural dimension may therefore be strongly observable and predictive in isolation while remaining largely redundant in the full feature set.

Rather than proposing a new predictor of IDS performance, this work contributes a controlled explanatory framework that separates residual structural magnitude from attack detectability and delineates the conditions under which their relationship is informative. Conventional IDS metrics quantify detection performance, whereas the structural observability profile, SLS, phase-specific analysis, and dimension-ablation analysis characterize the dimensional composition, temporal prevalence, inter-burst persistence, predictive utility, unique contribution, and redundancy of observable structural evidence. Accordingly, aggregate recall should be interpreted together with these properties when evaluating IDSs under encryption. This joint interpretation enables limited detectability to be associated more precisely with weak residual structural magnitude, low temporal prevalence, limited inter-burst persistence, redundancy, or factors not captured by marginal divergence. Practically, these insights may inform the design of resource-efficient edge IDSs by identifying potentially redundant feature dimensions and guide the evaluation of targeted traffic-padding strategies for highly observable protocol-lifecycle components.

\section{Conclusion and Future Work}
\label{sec:conclusion}

This paper presented an explanatory framework for analyzing residual structural observability and attack detectability in encrypted OPC~UA traffic collected from an industrial private~5G testbed. SLS summarized residual structural magnitude, while phase-specific and dimension-ablation analyses examined temporal prevalence, inter-burst persistence, predictive utility, unique contribution, and redundancy. 

SLS generally followed the within-family decline in recall but did not reproduce cross-family detectability ordering. It should therefore be interpreted as a first-order summary of residual structural magnitude that is informative for explaining detectability trends within the same attack mechanism, rather than as a universal predictor across heterogeneous mechanisms. Phase-specific analysis showed how observable evidence was distributed and persisted over an attack campaign, while dimension-ablation analysis clarified its predictive utility, unique contribution, and redundancy. Together, these analyses show that detectability depends not only on the magnitude of observable structural deviation, but also on its temporal distribution and predictive role. 

Future work will extend the analysis to additional attack mechanisms and explore multivariate temporal measures.

\section*{Acknowledgment}

This research is funded by dtec.bw -- Digitalization and Technology Research Center of the Bundeswehr. dtec.bw is funded by the European Union -- NextGenerationEU (project ``Digital Sensor-2-Cloud Campus Platform'' (DS2CCP), \url{https://dtecbw.de/home/forschung/hsu/projekt-ds2ccp}).

The authors would like to thank F. Mueller, J. Jockram, K. Singh, T. Kittel, ipoque GmbH, Deutsche Telekom, and Ericsson for their continuous support and valuable cooperation.
	
	\bibliographystyle{unsrtnat}
	\bibliography{references}
	
\end{document}